\documentclass[aps,prl,twocolumn,letterpaper,superscriptaddress,showpacs]{revtex4}
\usepackage{CJK}
\usepackage{keyval}%
\usepackage{graphicx}
\usepackage{dcolumn}
\usepackage{bm}
\usepackage{color}

\setkeys{Gin}{width=0.8\columnwidth,clip=true}

\newcommand{\ignore}[1]{}

\begin{document}
\begin{CJK*}{UTF8}{bsmi}

\title{One-Fe versus Two-Fe Brillouin Zone of Fe-Based Superconductors:\\
Creation of the Electron Pockets via Translational Symmetry Breaking}

\author{Chia-Hui Lin}
\affiliation{Condensed Matter Physics and Materials Science Department,
Brookhaven National Laboratory, Upton, New York 11973, USA}
\affiliation{Department of Physics and Astronomy, Stony Brook University, Stony Brook, New York 11794, USA}

\author{Tom Berlijn}
\affiliation{Condensed Matter Physics and Materials Science Department,
Brookhaven National Laboratory, Upton, New York 11973, USA}
\affiliation{Department of Physics and Astronomy, Stony Brook University, Stony Brook, New York 11794, USA}

\author{Limin Wang}
\affiliation{Condensed Matter Physics and Materials Science Department,
Brookhaven National Laboratory, Upton, New York 11973, USA}
\author{Chi-Cheng Lee}
\affiliation{Condensed Matter Physics and Materials Science Department,
Brookhaven National Laboratory, Upton, New York 11973, USA}

\author{Wei-Guo Yin}
\affiliation{Condensed Matter Physics and Materials Science Department,
Brookhaven National Laboratory, Upton, New York 11973, USA}

\author{Wei Ku}
\altaffiliation{corresponding email: weiku@bnl.gov}
\affiliation{Condensed Matter Physics and Materials Science Department,
Brookhaven National Laboratory, Upton, New York 11973, USA}
\affiliation{Department of Physics and Astronomy, Stony Brook University, Stony Brook, New York 11794, USA}

\date{\today}

\begin{abstract}
We investigate the physical effects of translational symmetry breaking in Fe-based high-temperature superconductors due to alternating anion positions.  In the representative parent compounds, including the newly discovered Fe-vacancy-ordered $\mathrm{K_{0.8}Fe_{1.6}Se_2}$, an unusual change of orbital character is found across the one-Fe Brillouin zone upon unfolding the first-principles band structure and Fermi surfaces, suggesting that covering a larger one-Fe Brillouin zone is necessary in experiments.  Most significantly, the electron pockets (critical to the magnetism and superconductivity) are found only created with the broken symmetry, advocating strongly its full inclusion in future studies, particularly on the debated nodal structures of the superconducting order parameter.
\end{abstract}

\maketitle
\end{CJK*}

One confusing/puzzling aspect of the new iron-based high-temperature superconductors is the dilemma of one-Fe vs. two-Fe description, concerning the translational symmetry of the system.  The generic crystal structure of these materials consists of two inequivalent Fe atoms, distinguished by the alternating tetrahedral coordination of the pnictogen or chalcogen anions (c.f.: Fig.~1a).  Since this coordination is known to impose dramatic impacts on the hopping integrals of Fe $d$ orbitals \cite{Lee}, the associated broken translational symmetry (from 1-Fe perspective) is expected to be physically significant and should be fully incorporated via the use of the unit cell that contains explicitly two Fe atoms.  Yet, the observed neutron scattering intensity \cite{Lumsden,Xu,Park:4503,Li:0503} shows little (if any) indication of such broken symmetry; it appears to follow simply the 1-Fe Brillouin zone (BZ) of a simple square lattice of Fe atoms (Fig.~1b).  Furthermore, out of convenience, most theoretical studies of superconductivity to date do not account fully for this broken translational symmetry, disregarding the rigorous symmetry constraint.  It is thus important and timely to clarify quantitatively various aspects of the effects from this broken symmetry (its relevance/irrelevance), and to settle, once for all, the confusing status of the field on the 1-Fe vs. 2-Fe perspective.

In this letter, three striking effects of the translational symmetry breaking potential (TSBP) are revealed by unfolding the \textit{ab initio} electronic band structures (EBSs) and Fermi surfaces (FSs) of representative parent compounds back to the 1-Fe BZ:  \textrm{i}) The folded Fe bands (``shadow'' bands) possess overall weak spectral weight, explaining the 1-Fe perspective advocated by the neutron measurements, and indicating the necessity of the larger 1-Fe BZ in angle resolved photoemission spectroscopy (ARPES) as well.  \textrm{ii}) The folding of the bands induces an unusual parity switching in their orbital characters, suggesting a change of photon polarization in ARPES.  \textrm{iii}) Most strikingly, the widely discussed electron Fermi pockets around $(\pi,0)$ and $(0,\pi)$ for supporting superconductivity would not have existed without the TSBP.  This advocates strongly the full inclusion of TSBP (the 2-Fe perspective) in theoretical understanding of superconductivity in these materials, and suggests a critical re-examination of the debated nodal structure of the superconducting order parameter \cite{Maier, Arita, Hanke, Chubukov, Gordon, Nakai, Hashimoto, DingEPL, Nakayama} on the electron pockets.

\begin{figure}[b]
     \centering
    \includegraphics[width=.8\columnwidth,clip=true]{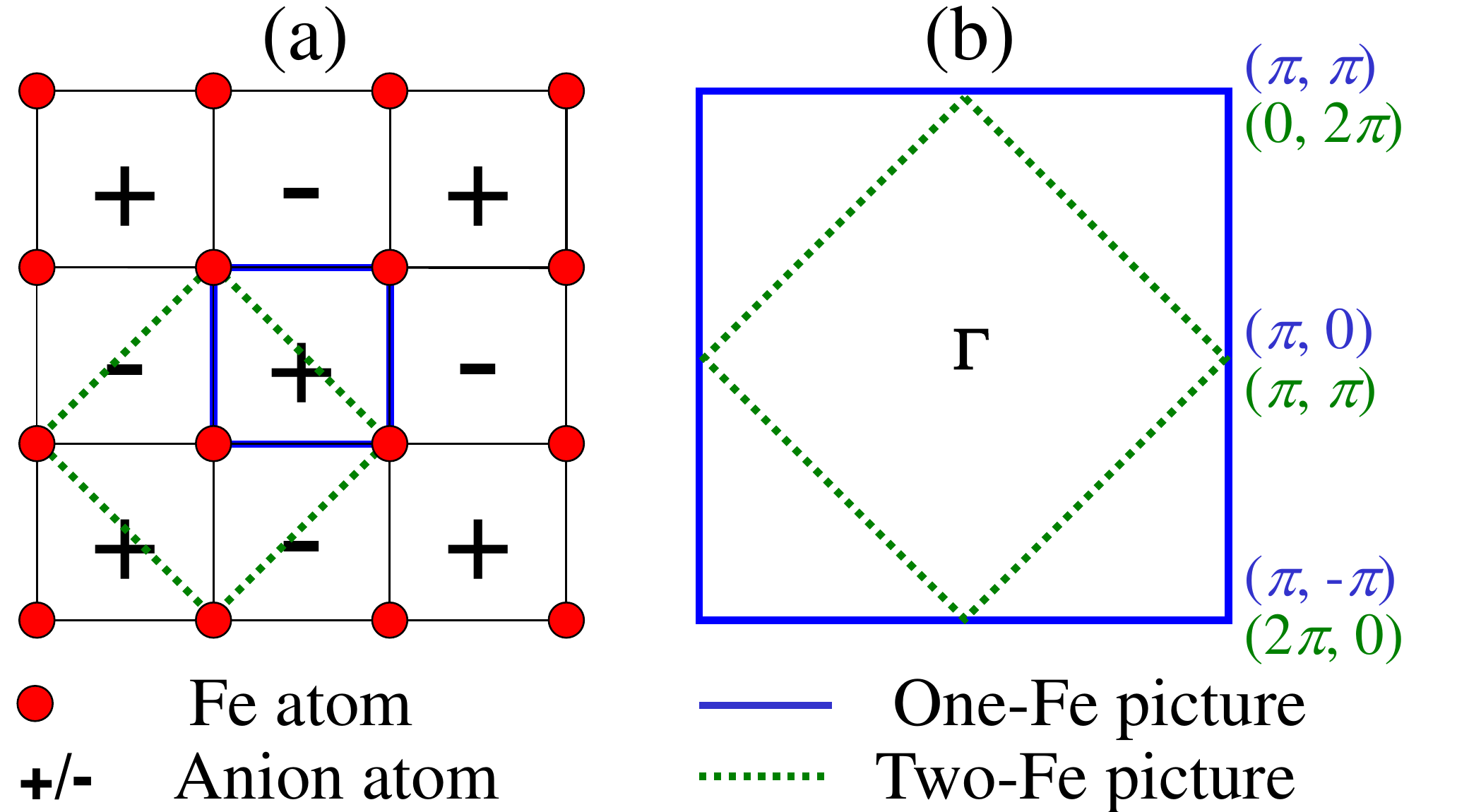}
   \caption{(Color online) Illustration of (a) one-Fe and two-Fe unit cells with +($-$) anions located above (below) the Fe plane, and (b) the corresponding first Brillouin zones.}
   \label{fig1}
\end{figure}

\begin{figure*}[t]
     \centering
    \includegraphics[width=2\columnwidth,clip=true]{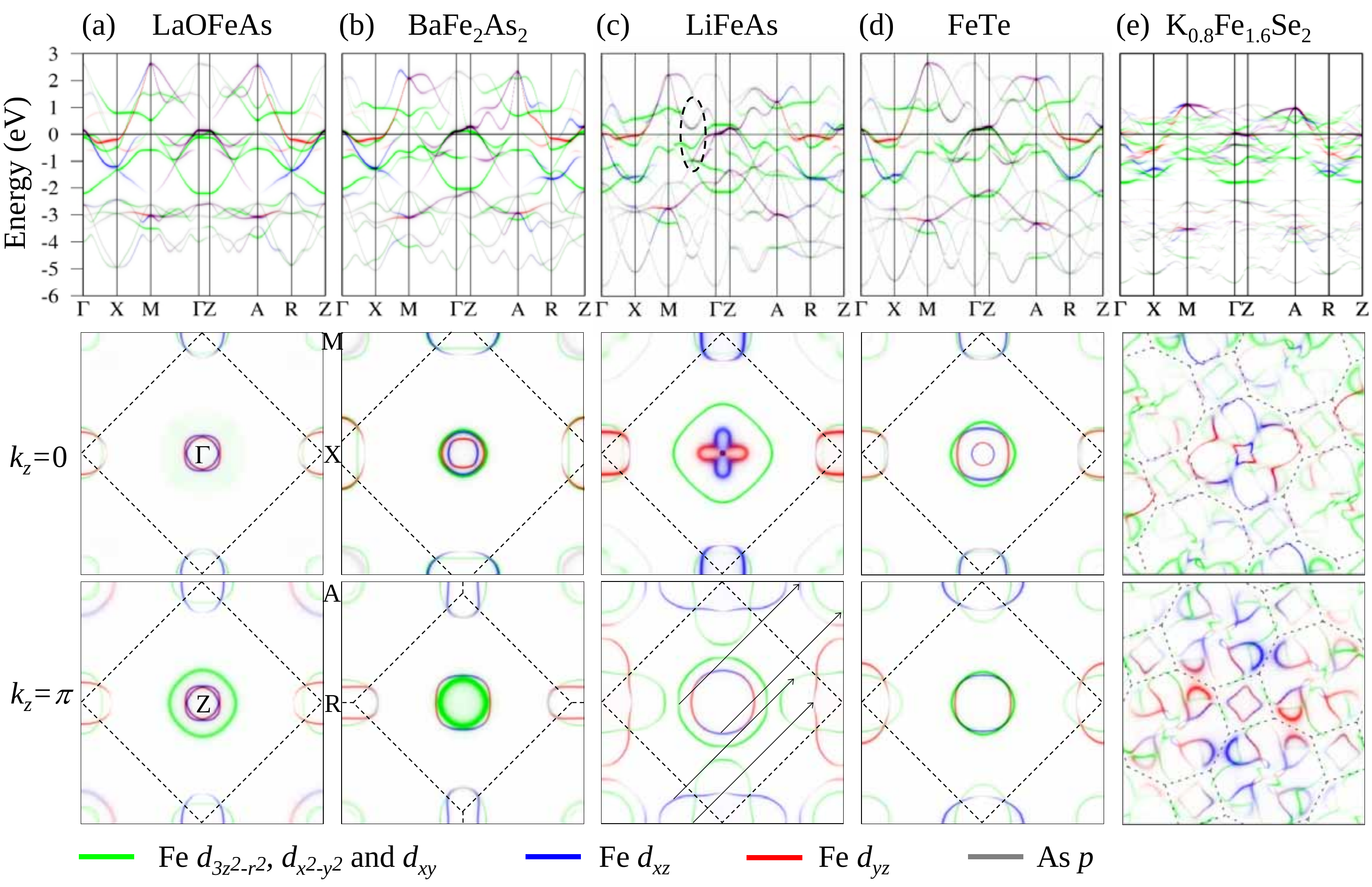}
   \caption{(Color online) Unfolded EBSs (top panel) and FSs at $k_z=0$ (middle panel) and $k_z=\pi$ (bottom panel) in 1-Fe BZ of (a) LaFeAsO, (b) $\mathrm{BaFe_2As_2}$, (c) LiFeAs, (d) FeTe, and (e) $\mathrm{K_{0.8}Fe_{1.6}Se_2}$ with spectral intensity colored by green (Fe $d_{3z^2 -r^2}$, $d_{x^2-y^2}$, and $d_{xy}$), blue (Fe $d_{xz}$), red (Fe $d_{yz}$) and grey (anion $p$).  The dashed lines marks the strict BZ boundaries. The arrows illustrate the folding of bands, and dashed ellipse in (c) points to an example of gap openings.}
   \label{fig2}
\end{figure*}

Our theoretical analysis is based on unfolded first-principles EBSs and FSs of the normal state in 1-Fe perspective, which reveals explicitly various aspects of the TSBP effects.  Standard density functional theory (DFT) calculations~\cite{Wien2k,supp} were conducted with local density approximation in the minimum unit cell (8 Fe for $\mathrm{K_{0.8}Fe_{1.6}Se_2}$ and 2 Fe for the rest).  Based on the DFT results, symmetry-respecting Wannier functions \cite{Ku2002} with Fe $d$ and anion $p$ characters were constructed to capture the low energy Hilbert space within [-6, 3] eV, based on which the low energy effective tight-binding Hamiltonians, $H$, were calculated.  Finally, unfolded EBSs and FSs were obtained via the recently developed first-principles unfolding method \cite{Ku:6401}.

The basic idea of our unfolding method~\cite{Ku:6401} is to simply represent the energy-, $\omega$-, dependent one-particle spectral function of the real systems (2-Fe zone) using the basis from a more symmetric reference system (1-Fe zone):
$A_{kn,kn}(\omega) = \sum_{KJ}|\langle kn| KJ\rangle|^2 A_{KJ,KJ}(\omega)$,
where $K$/$k$ denotes the crystal momentum of the original/reference system, $J$ the band index, and $n$ the Wannier orbital index.
This change of basis is made simple with the use of first-principles Wannier functions~\cite{Ku:6401}.  As demonstrated below, the unfolded EBSs and FSs provide explicit and detailed information on each band's coupling to the TSBP in an orbital specific manner.  Additionally, it can be shown \cite{Ku:6401} that the unfolded spectral function corresponds directly to the intensity of ARPES, as it includes the main matrix element effects of the measurement (expect the remaining atomic dipole matrix element to be determined based on the chosen photon polarization).  This use of ``regular'' momentum distinguishes our method from the glide symmetry-based unfolding employed by, for example, Andersen and Boeri \cite{Andersen:0008}, in which the twisted geometry does not have direct correspondence to the ARPES.  Similarly, the use of regular momentum is essential in the widely applied spin fluctuation studies \cite{Maier, Arita, Chubukov} of superconductivity via magnetic susceptibility, $\chi(q,\omega)$, since the momentum transfer $q$ concerns the \textit{difference} of two $k$ points.

Our resulting unfolded EBSs and FSs of the representative families in the nonmagnetic state are shown in Fig.~2, colored to emphasize the essential Fe $d$ orbitals. (Enlarged figures focusing around the Fermi energy are given in the supplementary materials~\cite{supp}.) A few generic features of unfolding can be clearly observed, for example, in Fig.~2c.   The most obvious one is the appearance of the shadow bands, generated from band ``folding'' via the TSBP.  Since here the TSBP is of momentum $q^{TSBP}=(\pi,\pi,0)$ in the 1-Fe BZ unit (except for $\mathrm{BaFe_2As_2}$ and $\mathrm{K_{0.8}Fe_{1.6}Se_2}$, whose double layer structure gives $q^{TSBP}=(\pi,\pi,\pi)$ instead) each band is folded from $k$ to $k+q^{TSBP}$, as illustrated by the arrows in the lower panel.  Note that the conservation of spectral weight dictates a weaker spectral weight for those ``main'' bands that develop stronger shadow bands.  Also associated with the shadow band formation are the additional gap openings occurring at the 2-Fe BZ boundaries, indicated by an ellipse in Fig.~2c as an example.  Obviously, the intensity of the shadow bands and gap opening size reflect (although not necessarily represent fully) the bands' coupling to the broken symmetry.

Fig.~2 shows clearly that the anion bands within [-6, -2] eV develop very strong shadows bands, of comparable intensity to the main bands.  This reflects their strong coupling to the TSBP, as it is the alternating positioning of the anion that breaks the translational symmetry.  Given that one can hardly distinguish the main bands from the shadow bands, it is obviously more convenient to consider these anion bands in the 2-Fe BZ.

In great contrast, the Fe bands near the Fermi level have rather weak shadow bands.  In fact, if it weren't for the gap openings (some of which are quite large), the Fe bands would have looked just like those from a simple 5-band system.  The overall weak intensity of shadow bands explains why neutron spectra appear to respect the 1-Fe BZ:  Even though the real symmetry of the system dictates the 2-Fe BZ, the folding of the spectrum is just not strong enough for a clear experimental identification. In fact, the lack of folded bands was also reported in a recent ARPES experiment. \cite{Brouet} Consequently, a larger 1-Fe BZ is $necessary$ in future ARPES measurements, since only about half of the EBSs are clearly observable in the 2-Fe BZ, where most ARPES to date were conducted.

Fig.~2 also reveals an interesting orbital-parity switching of the band folding, obvious from the change of color of the Fe-bands.  Consider the FSs, for example.  The blue ($d_{xz}$) and red ($d_{yz}$) bands of odd parity w.r.t. the Fe plane always have green ($d_{3z^2-r^2}$, $d_{x^2-y^2}$, and $d_{xy}$) shadow bands of even parity, and \textit{vice versa}.  This can be understood from the structure of TSBP in these systems.  Tab.~1 gives the nearest neighbor hopping integrals for the low-energy Fe $d$ bands in BaFe$_2$As$_2$ after integrating out As $p$ orbitals.  It shows that the alternating positioning of the anion (c.f. Fig.~1a) leads to an alternating sign of all $t_{even,odd}$, and thus breaks the translational symmetry.  Consequently, these terms form the main body of the TSBP, and dictate a switching of parity in the band character upon band folding.  This novel behavior is quite distinct from the common cases of ARPES, in which the replica of bands beyond the first BZ retain the orbital character.  Here, the weak replica always possess a different character across the BZ boundaries and thus require a different photon polarization to clearly observe, similar to the recent reports on Bi$_2$Sr$_2$CaCu$_2$O$_{8+\delta}$ \cite{Mans:7007}.

\begin{table}[t]
\caption{Nearest-neighbor hopping integrals (in eV) along the $x$-direction $\langle r\prime+(100),n|H|r\prime,n\prime\rangle$ among Fe $d$ Wannier orbitals for nonmagnetic $\mathrm{BaFe_2As_2}$.  The option in sign corresponds to two inequivalent Fe sites.  Bold font highlights orbitals with odd parity.}
\begin{ruledtabular}
\begin{tabular}{lccc|cc}
   & $3z^2-r^2$& $x^2-y^2$ & $xy$ & $\mathbf{xz}$ & $\mathbf{yz}$ \\
\hline
$3z^2-r^2$ & 0.03 & 0.31 &0 & 0 & $\mp0.12$    \\
$x^2-y^2$ & 0.31 & $-0.34$ &0 & 0 & $\pm0.34$    \\
 $xy$ & 0 &0 & $-0.12$ & $\mp0.22$ & 0    \\
 \hline
 $\mathbf{xz}$ & 0 & 0 &$\mp0.22$ & $-0.06$ &0 \\
 $\mathbf{yz}$ & $\mp0.12$ & $\pm0.34$ & 0 &0 & $-0.32$ \\
\end{tabular}
\end{ruledtabular}
\end{table}

\begin{figure}
     \centering
    \includegraphics[width=0.865\columnwidth,clip=true]{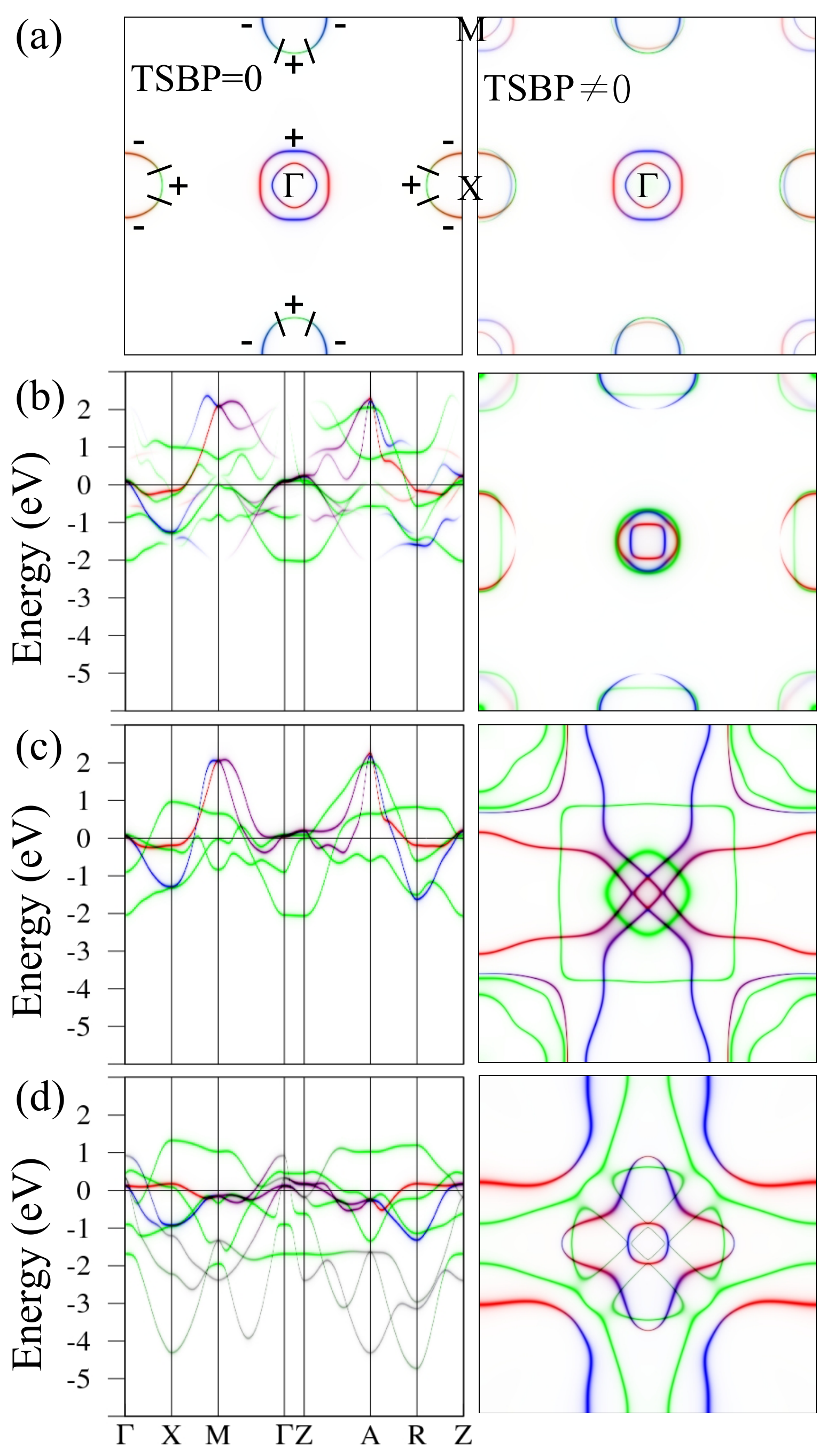}
   \caption{(Color online) (a) Folding of complete electron pockets (from Ref.~\cite{Maier}) and illustration of currently proposed nodal structure of the superconducting order parameter. (b) First-principles results showing $incomplete$ electron pockets instead.  (c)(d) Demonstration of loss of electron pockets by dropping TSBP in both 5- and 8-band VCA descriptions.}
   \label{fig3}
\end{figure}

The most significant feature revealed in Fig.~2 is the \emph{incompleteness} of the unfolded electron pockets around the X (and R) points.  Taking BaFe$_2$As$_2$ in Fig.~2b as an example, near the X=$(\pi,0,0)$ point, the intensity of the strong red pocket vanishes dramatically near the $\Gamma$-X path.  Consequently, only the green shadow pocket of $d_{xy}$ character, folded from the blue pocket around R=$(0,\pi,\pi)$ (c.f. bottom panel of Fig.~2b), is visible here, in agreement with recent ARPES measurement \cite{Ding}.  In fact, none of the unfolded pockets near X and R points are complete in Fig.~2, contrary to the common assumption that each X/R point has one strong $complete$ pocket and obtains a weaker shadow pocket via band folding (c.f. Fig.~3a).  All the electron pockets here are instead formed by $combining$ main bands near X and R points via the TSBP.  In other words, without breaking the 1-Fe translational symmetry, the essential electron pockets would have \textit{never existed} in these systems.

To better illustrate this important finding, let's construct a translational symmetric Hamiltonian (in 1-Fe unit) via the virtual crystal approximation (VCA) to the above effective Fe $d$-band Hamiltonian:
\begin{eqnarray}
H_{r,n;0,n\prime}^{VCA} = \sum_{r\prime}H_{r\prime+r,n;r\prime,n\prime}/\sum_{r\prime}1, \nonumber
\end{eqnarray}
where $r$ denotes the 1-Fe unit cell index.
This effectivley zeros out all TSBP (the above $t_{even,odd}$ terms), and keeps all translational symmetric terms intact.  The resulting EBS and FSs at $k_z=0$ are given in Fig.~3c.  Even though the overall EBS still follows the original structure in Fig.~3b (minus the shadow bands and gap opening obviously), the topology of the FSs is drastically modified.  In particular, there are no electron pockets around the X point anymore.  The same is found in the VCA of our original Hamiltonian containing Fe $d$ and As $p$ orbitals (Fig.~3d).  Evidently, the TSBP is instrumental in creating the electron pockets, and thus should be fully included in future theoretical modeling of magnetism and superconductivity.

Our findings have direct and significant implications on the heatedly debated issue of nodal structures of the superconducting order parameter on the electron pockets.  Current spin fluctuation theories \cite{Maier,Arita,Hanke} suggest accidental nodes in the $s_{+-}$ order parameter on the electron pockets (see Fig.~3a), due to strong inter-electron-pocket scattering.  While the existence of nodes appears to be supported by the interpretations of the penetration depth and several other measurements \cite{Gordon, Nakai, Hashimoto}, it contradicts with the nodeless and almost isotropic gaps observed in ARPES \cite{DingEPL, Nakayama}.  We find that precisely near the region of the nodes, ARPES would have negligible intensity, and thus can easily miss the nodal structure.  On the other hand, the above mentioned  theories did not incorporated appropriately the essential TSBP, and consequently are based on FSs of $qualitatively$ different spectral intensity and orbital structures.  Specifically, one would expect that the above incompleteness of the electron pockets, the mismatch in the orbital characters, and the addition of non-diagonal coupling between the pockets, can all affect quite strongly the inter-pocket scattering and alter the position or even the existence of the calculated accidental nodes on the electron pockets.  Thus, a careful re-examination of the theoretical prediction would be of great interest and importance.

Finally, let's consider the second TSBP introduced by ordered 20\% Fe vacancies in $\mathrm{K_{0.8}Fe_{1.6}Se_2}$ \cite{Zavalij:4882}.  The long period of the vacancy ordering, corresponding to a small $q^{TSBP}=\pm(\frac{3\pi}{5}, \frac{\pi}{5}, \pi)$ and $\pm(\frac{-\pi}{5}, \frac{3\pi}{5}, \pi)$, leads to a tiny BZ, making it difficult to compare standard DFT results \cite{Cao:1344} with the experiments.  Our unfolded EBS and FSs thus offer a direct comparison with experimental spectra, and provide a few theoretical insights.  Contrary to the above discussions, the Fe-vacancy induced TSBP causes a strong coherent scattering of the Fe bands, producing a larger number of shadow bands and strong gap openings all over the 1-Fe BZ.  Consequently, the overall band width of the Fe bands reduces by about 1/4, much more than the anion bands.  Not surprisingly, the resulting FSs are seriously reconstructed, showing little resemblance to the generic FSs of other cases, particularly lacking the nesting of the Fermi pockets.  This disagrees drastically with recent ARPES experiment, which reported well-defined Fermi pockets \cite{Zhang}.  Thus, the Fe-vacancy must order only weakly (or locally) in the measured samples, as observed recently by electron microscopy \cite{Wang}.  Similarly, now it seems more obvious that the recent experimental finding of enhanced superconductivity by promoting disorder of Fe-vacancies \cite{Wen} is mostly through the recovery of approximately nested Fermi pockets.

In conclusion, our first-principles unfolded EBSs and FSs reveals three key features of the translational symmetry breaking due to alternating anion positioning, in all families of the Fe-based superconductors.  First, the folded shadow bands have rather weak spectral weight.  This explains the apparent respect to the 1-Fe BZ in neutron measurements, and highlights the necessity of covering the larger 1-Fe BZ in future ARPES experiments.  Second, TSBP induced band folding changes the orbital character to those with opposite parity w.r.t. the Fe plane.  This unusual phenomenon suggests a change of photon polarization in ARPES experiment.  Finally and most significantly, the electron pockets, critical to most theories of magnetism and superconductivity of these materials, only form via the TSBP.  Thus, full inclusion of the broken translational symmetry (e.g. using 2-Fe unit cell) is essential in future theories, particularly on the debated issue of nodal structure of the superconducting order parameter on the electron pockets.

Work funded by the U S Department of Energy, Office of Basic Energy Sciences DE-AC02-98CH10886 and by DOE-CMCSN.

\begin{widetext}
{\Large\bf Supplementary information}

\vspace{10mm}

\section{Details of \textit{ab initio} calculation and band structures}
All \textit{ab initio} calculations were conducted by density functional theory with local density approximation in linearized augmented plane wave basis, which is implemented in WIEN2k package \cite{sWien2k}. For each parent compound, the lattice constants and atomic positions were obtained from the experimental data of the high temperature nonmagnetic state of  LaFeAsO \cite{s1111}, $\mathrm{BaFe_2As_2}$ \cite{s122}, LiFeAs \cite{s111}, FeTe \cite{s11}, and $\mathrm{K_{0.8}Fe_{1.6}Se_2}$ \cite{s245}. The essential parameters used in the calculation are summarized in Tab.~2. We followed the default settings of version 10.1 with $R_{mt}K_{max} = 7$ and $l_{max} = 10$ to reach convergence of the ground state density. Then the symmetry respecting Wannier functions \cite{sKu2002, sLee} of Fe $d$ and As $p$ orbitals were constructed within the low energy Hilbert space from -6 to 3 eV. Thus, the orbital-resolved spectral function in the 1-Fe Brillouin zone \cite{sKu2010} can be calculated from the effective tight-binding Hamiltonian on Wannier orbital basis by the unfolding method \cite{sKu2010}. Especially for the two-dimensional Fermi surfaces in $k_z$= 0 and $k_z$ = $\pi$ planes in the letter, we plotted the spectral function on the 400 x 400 $k$ point grid in order to capture all fine features. The unfolded band structures focusing on $\pm$2 eV energy around Fermi energy were plotted in Fig.~4 to reveal the details of Fe $d$ bands around Fermi energy.

\begin{table}[b]
\caption{Lattice constants and $k$ meshes used in the density functional theory calculation.}
\begin{ruledtabular}
\begin{tabular}{l|ccccc}
   Lattice constants& LaFeAsO& $\mathrm{BaFe_2As_2}$& LiFeAs & FeTe & $\mathrm{K_{0.8}Fe_{1.6}Se_2}$ \\
\hline
a=b (Bohr)  & 7.615732 & 7.488039  & 7.127100  & 7.221591 & 16.47577   \\
c (Bohr) & 16.51017 & 24.59819 & 12.01380 & 11.84764& 26.61737    \\
\hline
 $k$ mesh & $16\times16\times7$ & $13\times13\times13$ & $14\times14\times8$ & $17\times17\times10$ & $11\times11\times11$    \\
\end{tabular}
\end{ruledtabular}
\end{table}

\begin{figure}
    \centering
  \includegraphics[width=1\columnwidth,clip=true]{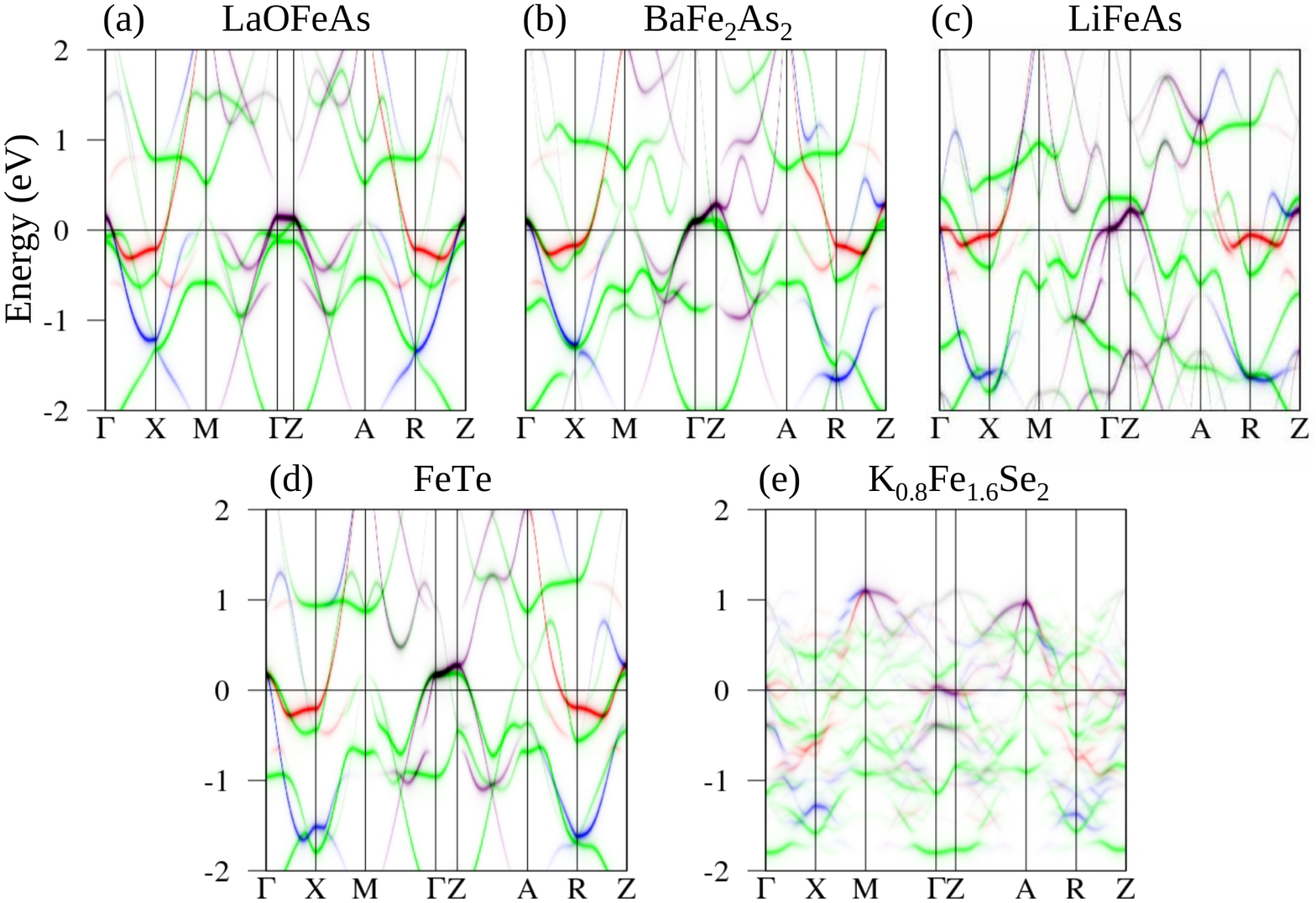}
  \caption{Unfolded electronic band structure in 1-Fe Brillouin zone of (a) LaFeAsO, (b) $\mathrm{BaFe_2As_2}$, (c) LiFeAs, (d) FeTe, and (e) $\mathrm{K_{0.8}Fe_{1.6}Se_2}$ with spectral intensity colored by green (Fe $d_{3z^2 -r^2}$, $d_{x^2-y^2}$, and $d_{xy}$), blue (Fe $d_{xz}$), red (Fe $d_{yz}$), and grey (anion $p$).}
  \label{fig:fig1}
\end{figure}

\section{Details of Virtual Crystal approximation}

In the presence symmetry breaking potential, a translationally symmetric Hamiltonian can be constructed via the virtual crystal approximation (VCA):
\begin{eqnarray}
H_{r,n;0,n\prime}^{VCA} = \sum_{r\prime}H_{r\prime+r,n;r\prime,n\prime}/\sum_{r\prime}1,
\end{eqnarray}
where $H_{r\prime+r,n;r\prime,n\prime}$ and $H_{r,n;0,n\prime}^{VCA}$ are Hamiltonians with broken translational symmetry in the basis of Wannier orbitals $n$ located at normal cell index $r$. The summation of $r\prime$ is over inequivalent normal cell lattice vectors in a single supercell (namely the positions of two or more inequivalent Fe sites). Thus, the symmetry breaking terms of the original Hamiltonian would be averaged out, but the symmetry respecting terms would be kept.

In the Fe-based superconductors, one can choose whether to integrate the As $p$ orbitals into Fe $d$ Wannier orbitals or not, during the construction of the Wannier orbitals. If As $p$ orbitals are integrated out, we will effectively obtain the five-band VCA Hamiltonian from Eq.~1. If not, the eight-band VCA Hamiltonian (5 Fe $d$ and 3 As $p$) will be obtained. Fig. 3 in the letter shows clearly that VCA has a larger impacts in the 8-band case.  This is easily understood from the following consideration.  In the 8-band case, the symmetry breaking terms are mostly those hopping terms involving Fe $d$ and As $p$ orbitals, since As $p$ orbitals are located in different locations.  Upon integrating out these terms, they effectively renormalize the remaining hopping terms between Fe $d$ orbitals, both the symmetry breaking and symmetry respecting ones.  By throwing out the symmetry breaking terms at the eight-band level via VCA, one also removes the renormalization to the symmetry respecting hopping between Fe $d$ orbitals as well.  In either case, however, the qualitative effect of losing the electron pockets persists, since it is mostly dictated by the symmetry property of the system and is thus more robust against detail changes of the Hamiltonian.

\end{widetext}


\begin{thebibliography}{00}
\bibitem{Lee}              C.-C. Lee \textit{et al}., Phys. Rev. Lett. {\bf103}, 267001 (2009).
\bibitem{Park:4503}         J. T. Park \textit{et al}., Phys. Rev. B {\bf82}, 134503 (2010).
\bibitem{Xu}                Z. Xu \textit{et al}., Phys. Rev. B {\bf82}, 104525 (2010).
\bibitem{Li:0503}           H.-F. Li \textit{et al}., Phys. Rev. B {\bf82}, 140503(R) (2010).
\bibitem{Lumsden}           M. D. Lumsden \textit{et al}., Nature Phys. {\bf6}, 182 (2010).

\bibitem{Maier}              S. Graser \textit{et al}., New J. Phys. {\bf11}, 025016 (2009).
\bibitem{Chubukov}          A. V. Chubukov \textit{et al.}, Phys. Rev. B {\bf78}, 134512 (2008).
\bibitem{Arita}             R. Arita and H. Ikeda, J. Phys. Soc. Jpn. {\bf78}, 113707 (2009).
\bibitem{Hanke}             R. Thomale \textit{et al.}, Phys. Rev. Lett. {\bf106}, 187003 (2011).
\bibitem{Gordon}            R. T. Gordon \textit{et al}., Phys. Rev. B {\bf79}, 100506(R) (2009).
\bibitem{Nakai}             Y. Nakai \textit{et al}., Phys. Rev. B {\bf81}, 020503 (2010).
\bibitem{Hashimoto}         K. Hashimoto \textit{et al}., Phys. Rev. B {\bf81}, 220501 (2010).
\bibitem{DingEPL}              H. Ding \textit{et al}., Eur. Phys. Lett., {\bf83} 47001 (2008).
\bibitem{Nakayama}          K. Nakayama \textit{et al}., Phys. Rev. Lett. {\bf105}, 197001 (2010).
\bibitem{Wien2k}            K. Schwarz \textit{et al}., Comput. Phys. Commun. {\bf147}, 71 (2002).
\bibitem{supp}              See supplementary information.
\bibitem{Ku2002}            W. Ku \textit{et al}., Phys. Rev. Lett. {\bf89}, 167204 (2002).
\bibitem{Ku:6401}           W. Ku \textit{et al}., Phys. Rev. Lett. {\bf 104}, 216401 (2010).
\bibitem{Andersen:0008}     O. K. Andersen and L. Boeri, Annalen der Physik {\bf 523}, 8 (2011).
\bibitem{Brouet}            V. Brouet \textit{et al}., arXiv:1105.5604 (2011).
\bibitem{Mans:7007}         A. Mans \textit{et al}., Phys. Rev. Lett. {\bf96}, 107007 (2006).

\bibitem{Ding}              H. Ding, private communication (2011).

\bibitem{Zavalij:4882}      P. Zavalij \textit{et al}., Phys. Rev. B {\bf83}, 132509 (2011).

\bibitem{Cao:1344}          C. Cao and J. Dai, arXiv:1102.1344 (2011).
\bibitem{Zhang}             Y. Zhang \textit{et al}., Nature Mater. {\bf10}, 273 (2011).
\bibitem{Wang}              Z. Wang \textit{et al}., Phys. Rev. B {\bf83}, 140505(R) (2011).
\bibitem{Wen}               F. Han \textit{et al}., arXiv:1103.1347 (2011).

\end{thebibliography}

\begin{thebibliography}{10}
\bibitem{sWien2k}           K. Schwarz \textit{et al}., Comput. Phys. Commun. {\bf147}, 71 (2002).
\bibitem{s1111}             C. de la Cruz \textit{et al}., Nature (London) 453, 899 (2008).
\bibitem{s122}              M. Rotter \textit{et al}., Phys. Rev. B {\bf78}, 020503(R) (2008).
\bibitem{s111}              X. Wang \textit{et al}., Solid State Commun. {\bf148}, 538 (2008).
\bibitem{s11}               D. M. Finlayson \textit{et al}.,Proc. Phys. Soc. B {\bf69} 860 (1956).
\bibitem{s245}              P. Zavalij \textit{et al}., Phys. Rev. B {\bf83}, 132509 (2011).
\bibitem{sKu2002}           W. Ku \textit{et al}., Phys. Rev. Lett. {\bf89}, 167204 (2002).
\bibitem{sLee}              C.-C. Lee \textit{et al}., Phys. Rev. Lett. {\bf103}, 267001 (2009).
\bibitem{sKu2010}           W. Ku \textit{et al}., Phys. Rev. Lett. {\bf 104}, 216401 (2010).

\end{thebibliography}
\end{document}